\documentclass{emulateapj}

\newcommand{\gray}{$\gamma$-ray}
\newcommand{\grays}{$\gamma$-rays}

\newcommand{\adv}{Adv.\ Space Res.}
\newcommand{\app}{APh}

\newcommand{\pubbook}[5]{#5, #1, #2 #3, #4}
\newcommand{\pubjournal}[5]{#4, #1, #2, #3}

\shorttitle{}
\shortauthors{Moskalenko et al.}

\journalinfo{\footnotesize \sc The Astrophysical Journal, 652: L65--L68, 2006 November 20}

\begin{document}

\title{Inverse Compton scattering on solar photons,\\
heliospheric modulation, and neutrino astrophysics}

\author{Igor V. Moskalenko\altaffilmark{1}}
\affil{
   Hansen Experimental Physics Laboratory, 
   Stanford University, Stanford, CA 94305
\email{imos@stanford.edu}}
\altaffiltext{1}{Also Kavli Institute for Particle Astrophysics and Cosmology,
Stanford University, Stanford, CA 94309}

\author{Troy A. Porter}
\affil{
  Santa Cruz Institute for Particle Physics,
  University of California, Santa Cruz, CA 95064
\email{tporter@scipp.ucsc.edu}}

\and

\author{Seth W. Digel$^1$}
\affil{Stanford Linear Accelerator Center, 
2575 Sand Hill Road, Menlo Park, CA 94025
\email{digel@stanford.edu}}

\begin{abstract}

We study the inverse Compton scattering of solar photons by  Galactic
cosmic-ray electrons.  We show that the \gray\ emission from this
process is substantial with the  maximum flux in the direction of the
Sun; the angular distribution of the emission is broad.  This
previously-neglected foreground should be taken into account in
studies of the diffuse Galactic and extragalactic \gray\ emission.
Furthermore, observations  by GLAST can be used to monitor the
heliosphere and determine the  electron spectrum as a function of
position from  distances as large as Saturn's orbit to close proximity
of the Sun, thus enabling unique studies of solar modulation.  This
paves the way for the determination of other Galactic cosmic-ray
species,  primarily protons, near the solar surface which will lead to
accurate  predictions  of \grays\ from $pp$-interactions in the solar
atmosphere.  These albedo \grays\ will be observable by GLAST,
allowing the study of deep atmospheric layers,  magnetic field(s), and
cosmic-ray cascade development.  The latter is necessary to calculate
the neutrino flux from $pp$-interactions at higher energies ($>$1
TeV).  Although this flux is small,  it is a ``guaranteed flux'' in
contrast to other astrophysical sources of neutrinos, and may be
detectable by km$^3$ neutrino telescopes of the near  future, such as
IceCube.  Since the solar core is opaque for very high-energy
neutrinos, directly studying the mass distribution of the solar core
may thus be  possible.

\end{abstract}


\section{Introduction}
Interactions of Galactic cosmic-ray (CR) nuclei with the solar atmosphere 
have been predicted to be a source of very high energy (VHE) neutrinos 
\citep{Moskalenko1991,Seckel1991,Ingelman1996}, and
\grays\ \citep{Seckel1991}. 
They are the decay products of
charged and neutral pions produced in interactions
of CR nucleons with gas. 
The predictions
for these albedo \grays\ give an integral flux
$F_\gamma(>100\ {\rm MeV})\sim(0.2-0.7)\times10^{-7}$ cm$^{-2}$ s$^{-1}$,
while analysis of EGRET \gray\ telescope data has yielded only the upper limit
$F_\gamma(>100\ {\rm MeV})=2.0\times10^{-7}$ cm$^{-2}$ s$^{-1}$
\citep{Thompson1997}.
At lower energies ($<$100 MeV) a contribution from CR electron
bremsstrahlung in the solar atmosphere
may exceed that from 
$\pi^0$-decay \citep[see Fig.~4 in][]{Strong2004}.

Cosmic-ray electrons comprise $\sim$1\% of the total CR flux.
However, they propagate over the heliospheric volume, which is large compared
to the solar atmosphere where the albedo \grays\ are produced.
The Sun emits photons that are targets for inverse Compton (IC) scattering
by CR electrons.
As a result the heliosphere is a diffuse source of \grays\ with a 
broad angular distribution.
In this paper, we evaluate the importance of IC scattering within the 
heliosphere, and discuss the consequences of its measurement by such 
instruments as the upcoming Gamma Ray Large Area Space Telescope (GLAST) 
mission. In the following, we use units $\hbar=c=m_e=1$.

\section{Anisotropic inverse Compton scattering in the heliosphere}
\begin{figure}
\centerline{\includegraphics[width=3in]{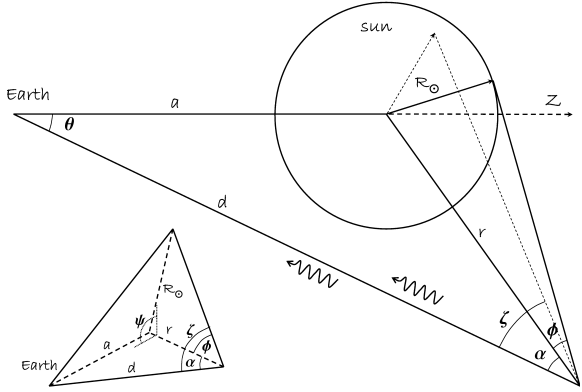}}
\caption{Variables involved in the calculation. The large plot shows a
2-dimensional sketch; the dotted lines are a projection of the 3-dimensional
angles. The inset defines the angles in 3 dimensions.}
\label{fig1}
\vspace{-0.10in}
\end{figure}

The distribution of CR electrons within the heliosphere is approximately
isotropic.
However, the distribution of solar photons is distinctly anisotropic, with 
the photons propagating outward from the Sun.
The expression for the IC production spectrum for an 
arbitrary photon angular distribution is given by eq.~(8) from 
\citet{Moskalenko2000}:
\begin{eqnarray}
\label{IC.8}
&&\frac{dR(\gamma_e,\epsilon_1)}{d\epsilon_2} = 
   \frac{\pi r_e^2}{\epsilon_1(\gamma_e-\epsilon_2)^2}
   \int_{\Omega_\nu} d\Omega_\nu\, Q_\nu(\Omega_\nu)\\
&&\quad\times
   \left[2 -2\frac{\epsilon_2}{\gamma_e}\left(\frac{1}{\epsilon'_1}+2\right)
   +\frac{\epsilon_2^2}{\gamma_e^2}\left(\frac{1}{{\epsilon'_1}^2}
      +2\frac{1}{\epsilon'_1}+3\right)
   -\frac{\epsilon_2^3}{\gamma_e^3}\right],\nonumber
\end{eqnarray}
where $r_e$ is the classical electron radius, 
$\gamma_e$ is the electron Lorentz-factor, 
$\epsilon_1$ and $\epsilon'_1$ are the energies of the background photon
in the laboratory system (LS) and electron rest system, correspondingly,  
$\epsilon_2$ is the LS energy of upscattered photon, 
$
\epsilon_2 \le 2\gamma_e \epsilon'_1/ (1+2\epsilon'_1), 
\epsilon'_1 = \epsilon_1\gamma_e(1+\beta_e\cos\zeta),
$
$\zeta$ is the LS angle between the momenta of the electron and
incoming photon ($\zeta=0$ for head-on collisions, see
Fig.~\ref{fig1}),
$
\epsilon_{2\max}= 4\epsilon_1\gamma_e^2/(1+4\epsilon_1\gamma_e)
$
is the maximal energy of the upscattered photons, and 
$Q_\nu(\Omega_\nu)$ is the normalised angular distribution of target 
photons at a particular spatial point ($\int d\Omega_\nu\, Q_\nu = 1$).

\placefigure{fig1}

The \gray\ flux as a function of $\theta$ (see Fig.~\ref{fig1}) can be
calculated by integrating eq.~(\ref{IC.8}) along the line-of-sight taking
into account the distribution of electrons and solar photons:
\begin{eqnarray}
\label{flux}
\frac{dF_\gamma}{d\epsilon_2}=
\frac14\int_L dx\,\frac{R_\sun^2}{r^2}\,
&& \int d\gamma_e\, \frac{dJ_e(r,\gamma_e)}{d\gamma_e}\\ 
&&\times
\int d\epsilon_1 \frac{dn_{bb}(\epsilon_1, T_\sun)}{d\epsilon_1}\,
\frac{dR(\gamma_e,\epsilon_1)}{d\epsilon_2},\nonumber
\end{eqnarray}

\noindent
where the factor $1/4$ comes from the angular distribution of 
photons for the case of an emitting surface 
(see \citealt{Moskalenko2000}, eqs.~[24]-[26]),
$J_e(r, \gamma_e)$ is the
electron intensity, $r$ is the radial distance from the Sun, 
$R_\sun$ is the solar radius, and $dn_{bb} (\epsilon_1, T_\sun)/d\epsilon_1$ is 
a blackbody distribution at temperature $T_\sun$.

Close to the Sun, $Q_\nu (\Omega_\nu)$ is given by
\begin{eqnarray}
&&Q_\nu(r,\phi) = \frac1{\pi P(r)}\left(1-\frac{r^2}{R_\sun^2}\sin^2\phi \right)^{1/2},\\
&&P(r) = 1-\frac{r^2-R_\sun^2}{2rR_\sun}\ln\left(\frac{r+R_\sun}{r-R_\sun}\right),\\
&&\cos\zeta = \cos\alpha \cos\phi +\sin\alpha \sin\phi \cos\psi,
\end{eqnarray}
where 
$
\int_0^{2 \pi} d\psi \int^1_{\cos\phi^{\max}} Q_\nu(r,\phi)\, d \cos\phi = 1,
$
$\sin\phi^{\max}=R_\sun/r$. 
For sufficiently large distances, this reduces to a delta function, 
$Q_\nu(\theta)=\delta(\cos\zeta-\cos\alpha)$, where 
$\cos\alpha=(d-a\cos\theta)/(a^2+d^2-2ad\cos\theta)^{1/2}$, $a=1$ AU,
and the variables are illustrated in Fig.~\ref{fig1}.

\begin{figure}
\centerline{
\includegraphics[width=3.5in]{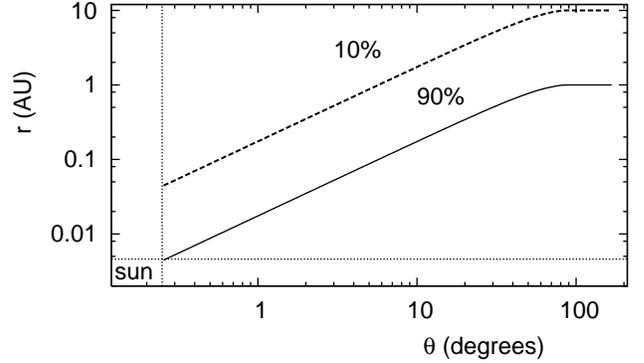}}
\caption{The range of heliospheric distances probed by the IC \grays\ versus
angular distance from the Sun. 90\% of the predicted \gray\ flux is produced
in the region between the solid and dashed lines.
Dotted lines show the solar size.}
\label{fig2}
\vspace{-0.07in}
\end{figure}

The IC \gray\ flux (eq.~[\ref{flux}]) essentially falls as $1/r$ with 
heliocentric distance.
Therefore, it is straightforward to 
estimate the region which contributes most to the IC emission. 
The inner boundary (in AU) is given by 
$$
r_1=\left\{
\begin{array}{ll}
\sin\theta,& \theta<90^\circ,\\
1,         & \theta\ge90^\circ.
\end{array}
\right.
$$
Approximately 90\% of the total \gray\ flux is produced between $r_1$ and 
$r_2=10r_1$ (Figure~\ref{fig2}).

The electron spectrum as a function of position in the heliosphere is obtained
by adopting the force-field approximation and deriving the radial
dependence of the modulation potential using a toy model.
This should be sufficient, as we seek only to illustrate the effect, and
are interested in electron energies above $\sim$5 GeV where the heliospheric 
modulation is moderate.
Assuming a separability
of the heliospheric diffusion coefficient 
$\kappa=\beta\kappa_1(r,t)\kappa_2(\gamma,t)$,
the modulation potential can be obtained \citep{forcefield}:
\begin{equation}
\Phi(r,t)= \int_r^{r_b(t)} dx\, \frac{V(x,t)}{3\kappa_1(x,t)},
\label{eq_Phi}
\end{equation}
where $t$ is the time variable, $r_b$ is the heliospheric boundary, 
and $V$ is the solar wind speed.

\citet{fujii05} have analysed the radial dependence of the mean free path,
$\lambda$, during the solar minima of Cycles 20/22 and Cycle 21 using the 
data from the IMP 8, Voyagers 1/2, and Pioneer 10 spacecraft. 
For Cycles 20/22 they obtain a parameterisation
$\lambda \propto r^{1.4}$ in the outer heliosphere and $\lambda \propto r^{1.1}$ 
in the inner heliosphere, where the break is at 10--15 AU.
For Cycle 21, an $r^{1.1}$ dependence fits well for both the inner 
and outer heliosphere.
Using these parameterisations, $\kappa_1\propto \lambda$, and 
assuming $V=const$ yields (Cycles 20/22) 
\begin{equation}
\Phi_1(r)= \frac{\Phi_0}{1.88}\left\{
\begin{array}{ll}
r^{-0.4}-r_b^{-0.4}, & r>r_0,\\
0.24+8(r^{-0.1}-r_0^{-0.1}), & r<r_0, 
\end{array}
\right.
\label{phi1}
\end{equation}
where $\Phi_0$ is the modulation potential at 1 AU, $r_0=10$ AU, $r_b=100$ AU,
and we have neglected the time dependence. 
For Cycle 21 we have 
\begin{equation}
\Phi_2(r)= \Phi_0 (r^{-0.1}-r_b^{-0.1})/(1-r_b^{-0.1}).
\label{phi2}
\end{equation}

\begin{figure}
\centerline{
\includegraphics[width=3.5in]{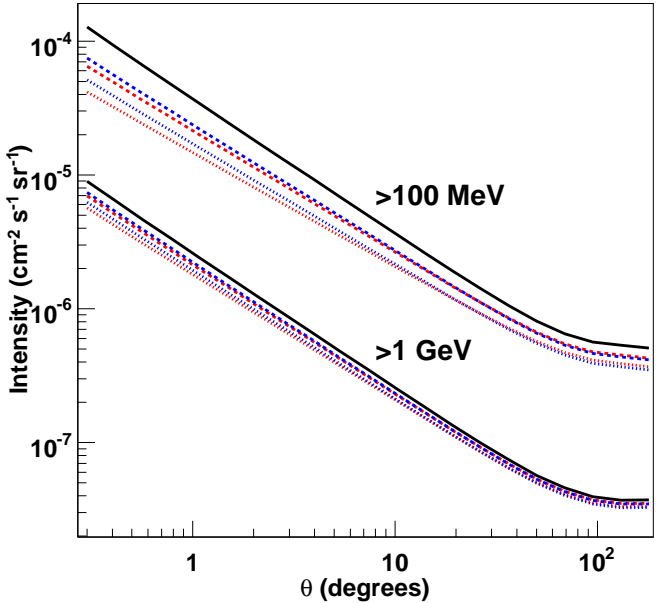}}
\caption{Integral intensity above 100 MeV and 1 GeV.
Black line: no modulation;
red lines: potential $\Phi_1$;
blue lines: potential $\Phi_2$.
Dashed lines: $\Phi_0 = 500$ MV; dotted lines: $\Phi_0 = 1000$ MV.}
\label{fig3}
\vspace{-0.07in}
\end{figure}

\placefigure{fig2}
\placefigure{fig3}

\section{Calculations}
The local interstellar CR electron spectrum is approximated by
$dJ_e(r_b,\epsilon_e)/d\epsilon_e=
2\times10^5(\epsilon_e+400)^{-3.22}$ cm$^{-2}$ s$^{-1}$ 
sr$^{-1}$ MeV$^{-1}$, where $\epsilon_e$ is the electron energy in MeV.
This expression matches
the interstellar local electron spectrum above $\sim$500 MeV 
as calculated by GALPROP in plain diffusion and reacceleration 
models \citep{Ptuskin2006}. 
The solar photospheric temperature 
is taken as $T=5770$ K \citep{Cox_book}. 

Figure \ref{fig3} shows the integral intensity above 100 MeV
and 1 GeV versus the angular distance from the Sun for no modulation,
and for modulated electron spectra according to eqs.~(\ref{phi1}) and 
(\ref{phi2}) with two modulation levels 
$\Phi_0=500$, 1000 MV; these correspond approximately to the 
solar minimum and maximum conditions, respectively.
The dependency on $\Phi_0$ is strong and thus the 
emission will vary over the solar cycle while the dependency on
the modulation potential parameterisation is much weaker.

The $68\%$ containment radius of the EGRET point spread function is 
$\sim 6^\circ$ at 100 MeV.
For $\theta < 6^\circ$, we calculate an integral flux
$F_\gamma(>100\ {\rm MeV})$ 
$\sim 1.6 \times 10^{-7}$ cm$^{-2}$ s$^{-1}$ (modulated).
The sum of the IC and albedo emission 
\citep{Seckel1991} over this region is consistent
with the EGRET upper limit of 
$2\times 10^{-7}$ cm$^{-2}$ s$^{-1}$ \citep{Thompson1997}.

\begin{figure}
\centerline{
\includegraphics[width=3.4in]{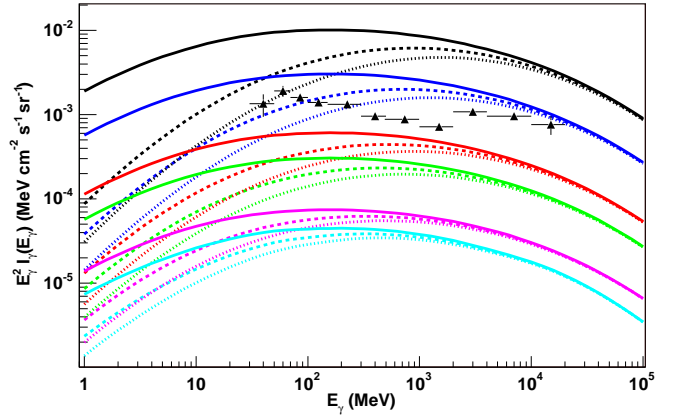}}
\caption{Differential intensities for selected $\theta$.
Line-sets (top to bottom): $0.3^\circ$, 
1$^\circ$, 5$^\circ$, 10$^\circ$, 45$^\circ$, and 180$^\circ$. 
Solid line: no modulation; dashed line: $\Phi_0 = 500$ MV; 
dotted line: $\Phi_0 = 1000$ MV.
Data points: diffuse extragalactic \gray\ flux \citep{Strong2004}.}
\label{fig4}
\vspace{-0.10in}
\end{figure}

The differential intensities for selected angles $\theta$ are
shown in Fig.~\ref{fig4}.  
The parameterisations $\Phi_{1,2}$, eqs.~(\ref{phi1}) and (\ref{phi2}), 
yield very similar differential \gray\ intensities; we thus show 
the results for $\Phi_1$ only.
Also shown is the extragalactic \gray\ background (EGRB)
obtained by \citet{Strong2004}, which includes true diffuse emission and 
an unresolved source component.
For $\theta < 5^\circ$, the solar IC emission is 
more intense than the EGRB. The latter is the major component of
the diffuse \gray\ emission outside of the Galactic plane \citep{Strong2004b}.
Table \ref{table1} gives the integral intensities above 10 MeV, 100 MeV, and 
1 GeV averaged over the whole sky. 
The values for 500 MV and 1000 MV are obtained by averaging over the 
fluxes for $\Phi_1$ and $\Phi_2$.
Above 100 MeV, the sky-averaged intensity is $\sim$5-10\% of the EGRB
obtained by \citet{Strong2004}.

Note that unresolved blazars are thought to be the major contributors 
to the EGRB \citep{stecker96,Kneiske2005}; thousands of them
will be resolved by GLAST. 
Therefore, the \citet{Strong2004}
EGRB flux is an upper limit for the true diffuse emission.

\section{Perspectives for GLAST}
The Large Area Telescope (LAT) high-energy \gray\ detector, 
in preparation for launch by NASA in late 2007, will have unprecedented 
angular resolution, effective area, and field of view for high-energy 
\grays\ \citep{McEnery2004}.  
The LAT will scan the sky continuously and provide essentially complete
sky coverage every 
2 orbits (approximately 3 hours).  
As a consequence, the light curve for any detectable source will be 
well sampled.  
Also, the scanning motion will permit frequent evaluation of the 
stability of the performance of the LAT with respect to ``standard
candle'' sources.
Based on the expected sensitivity of the 
LAT\footnote{http://glast.stanford.edu},
a source with flux $\sim$$10^{-7}$ cm$^{-2}$ s$^{-1}$ and the
hardness of the solar IC emission will be detectable on a daily basis when
the Sun is not close to the Galactic plane, where the diffuse emission is
brightest.  
Sensitive variability studies of the bright core of the IC
emission surrounding the Sun should be possible on weekly time scales.
With exposure accumulated over several months, the Sun should be resolved
as an extended source and potentially its IC, $\pi^0$-decay, and 
bremsstrahlung components separated spatially.

The extended IC emission is not isotropic,
but accumulated over
year timescales will be uniform in ecliptic longitude and brightest at low
ecliptic latitudes.  
The ecliptic plane crosses the Galactic plane near
the Galactic Centre, and the solar IC component may be important 
for investigating the nature of the reported ``halo'' of \grays\
about the Galactic Centre \citep{dixon1998}.

The solar IC contribution to the overall celestial diffuse emission can be
modelled directly as we have done here.  
Measurement of the spectrum of
solar IC emission in the near Sun direction fixes the emission at all
angles. 
This is especially true at high energies where the solar
modulation in the outer heliosphere is negligible.
Also, for any particular direction
of the sky the Galactic and extragalactic components can be derived from
the data by measuring variations in diffuse intensity with solar
elongation angle.

\begin{deluxetable}{lccc}
\tablecolumns{4}
\tablewidth{0pc}
\tabletypesize{\footnotesize}
\tablecaption{All-sky average integral intensity \label{table1}}
\tablehead{
\colhead{$E$} &
\colhead{$\Phi_0= 0$} &
\colhead{$500$ MV} &
\colhead{$1000$ MV} 
}
\startdata
$>$10  MeV & 5.9 & 3.7 & 2.5 
\\
$>$100 MeV & 0.7 & 0.6 & 0.5 
\\
$>$1   GeV & 0.05 & 0.05 & 0.05 
\enddata
\tablecomments{Units $10^{-6}$ cm$^{-2}$ s$^{-1}$ sr$^{-1}$.}
\end{deluxetable}

\section{Discussion and Conclusion}
In this paper we have studied the IC scattering of solar photons by
CR electrons\footnote{When this work
had already been completed
we learned about work by \citet{orlando2006}
on the same subject.}.
We have shown that the emission is significant and broadly distributed
with maximum brightness in the direction of the Sun.
The whole sky is shining in \grays\ contributing to
a foreground that would otherwise be ascribed to
the Galactic and extragalactic diffuse emission.

The IC emission from CR electrons depends on their spectrum
in the heliosphere and varies with the modulation level. 
Observations in different
directions can be used to determine the electron
spectrum at different heliocentric distances.
A sensitive \gray\ telescope
on orbit could monitor the heliosphere, providing
information on its dynamics. 
Such observations 
also could be used
to study the electron
spectrum in close proximity to the Sun, 
unreachable for direct measurements by spacecraft.
The assumed isotropy of the electron distribution everywhere
in the heliosphere is an approximation. 
Closer to the Sun,
the magnetic field and non-isotropic solar wind speed affect the
CR electron spectrum and angular distribution,
which in turn will produce asymmetries in the IC emission.
Observations of such asymmetries may provide us with information
about the magnetic field and solar wind speed at different
heliolatitudes including far away from the ecliptic.

What are the immediate implications?
Since solar modulation theory is well developed 
\citep[e.g.,][]{Zank2003,Ferreira2004},
accurate measurement of the electron spectrum near the solar 
surface will open the way to derive the spectra of other 
Galactic CR species, primarily protons, near the Sun. 
The CR proton spectrum
is the input to calculations
of \grays\ from $pp$-interactions in the solar atmosphere.
These predictions can be further tested using GLAST observations of 
pionic \grays\ from the Sun. In turn, this would 
provide information about the density profile of the solar atmosphere, 
magnetic field(s), and CR cascade development \citep{Seckel1991}.
The higher the energy of \grays, the higher the energy of the ambient
particles, and thus the depth of the layers tested is increased.
In conjunction with other solar monitors
this can bring understanding of the deep atmospheric 
layers, Sun spots, magnetic storms, and other solar activity.

Furthermore, understanding 
the solar atmosphere is necessary to calculate the neutrino flux
from $pp$-interactions at higher energies ($>$1 TeV), as the
CR shower development depends on the density distribution
and the underlying magnetic field structure
(see \citealt{Seckel1991} and \citealt{Ingelman1996} for details).
Although small, 
the VHE neutrino flux from the Sun is a ``guaranteed
flux'' in contrast to other astrophysical sources of neutrinos.
It may be detectable by km$^3$ neutrino telescopes of the near future, 
such as IceCube \citep[e.g.,][]{Hettlage2000,Ahrens2004},
on a timescale of a year or several years.
Therefore, only long term
periodicities associated with the solar activity could be detectable.
Since the solar core is opaque for VHE neutrinos 
\citep{Seckel1991,Moskalenko1993,Ingelman1996}, observations 
of the neutrino flux may provide us with information 
about the solar mass distribution.

To summarise, we have shown that the observation of the IC emission
from CR electrons in the heliosphere and its distribution on the
sky will open a new chapter in astrophysics. 
This makes a sensitive \gray\
telescope a useful tool to monitor the heliosphere and
heliospheric propagation of CR, and to study CR interactions in the 
solar atmosphere.

\acknowledgments
I.\ V.\ M.\ acknowledges partial support from NASA
Astronomy and Physics Research and Analysis Program (APRA) grant.
T.\ A.\ P.\ acknowledges partial support from the US Department of Energy.

\newpage
\noindent
{\footnotesize \sc The Astrophysical Journal, 664: L143, 2007 August 1}
\bigskip

\begin{center}
ERRATUM: ``INVERSE COMPTON SCATTERING ON SOLAR PHOTONS, HELIOSPHERIC MODULATION, \\
AND NEUTRINO ASTROPHYSICS'' (ApJ, 652, L65 [2006])

\medskip
\sc Igor V.\ Moskalenko, Troy A.\ Porter, Seth W.\ Digel
\end{center}


\medskip

\setcounter{figure}{2}
\setcounter{table}{0}
\setcounter{equation}{0}

We noticed an error in the description of the distribution of solar 
photons at an arbitrary distance from the Sun, equations (3) and (4). 
The correct expression is
\begin{eqnarray}
&& Q_\nu(r,\phi) = \frac1{2\pi}\left[1-\left(1-
\frac{R_\odot^2}{r^2}\right)^{1/2}\right]^{-1},\\
&& \left(1-\frac{R_\odot^2}{r^2}\right)^{1/2}\le\cos\phi\le1,
\label{eq1}
\end{eqnarray}
i.e.\ $Q_\nu(r,\phi)$ is independent of $\phi$ within the
solid angle covered by the Sun. 
Applying the correct angular distribution does not give results that are 
noticeably different from those 
obtained with the delta-function (pure radial) photon distribution.
Indeed, it should be the case
since in the energy range under consideration $\gamma_e\gg1$ and
the ambient photon angular distribution can be approximated by the 
delta-function.

We also discovered a numerical error in the code which affects the
results below $\sim$1 GeV, especially in case of small $\theta$. 
Figures \ref{fig3e} and 
\ref{fig4e} show the corrected integral and differential intensities.
Table 1 shows the corrected all-sky average integral intensities.
The $68\%$ containment radius of the EGRET point spread function is
$\sim$$6^\circ$ at 100 MeV.
For $\theta < 6^\circ$, the corrected integral flux is
$F_\gamma(>100\ {\rm MeV})$
$\sim (2.0-4.3) \times 10^{-7}$ cm$^{-2}$ s$^{-1}$, where the given range
corresponds to different modulation levels ($\Phi_0=1000-500$ MV).

\placefigure{fig3}
\placefigure{fig4}
\placetable{table1}

\newcommand{\wcap}{0.476\textwidth}

\begin{figure*}[h]
\centerline{
\includegraphics[width=3.6in]{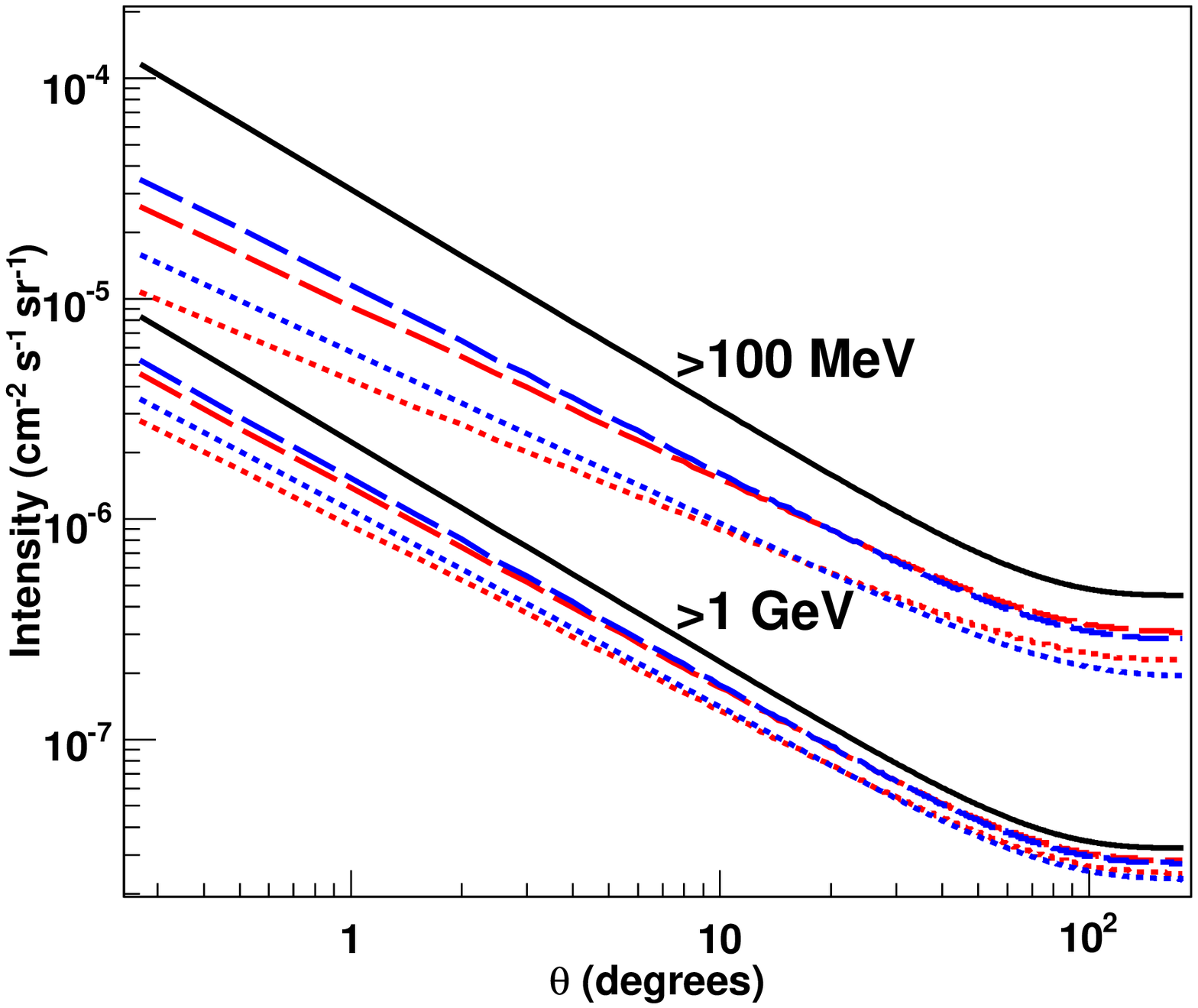}\hfill
\includegraphics[width=3.6in]{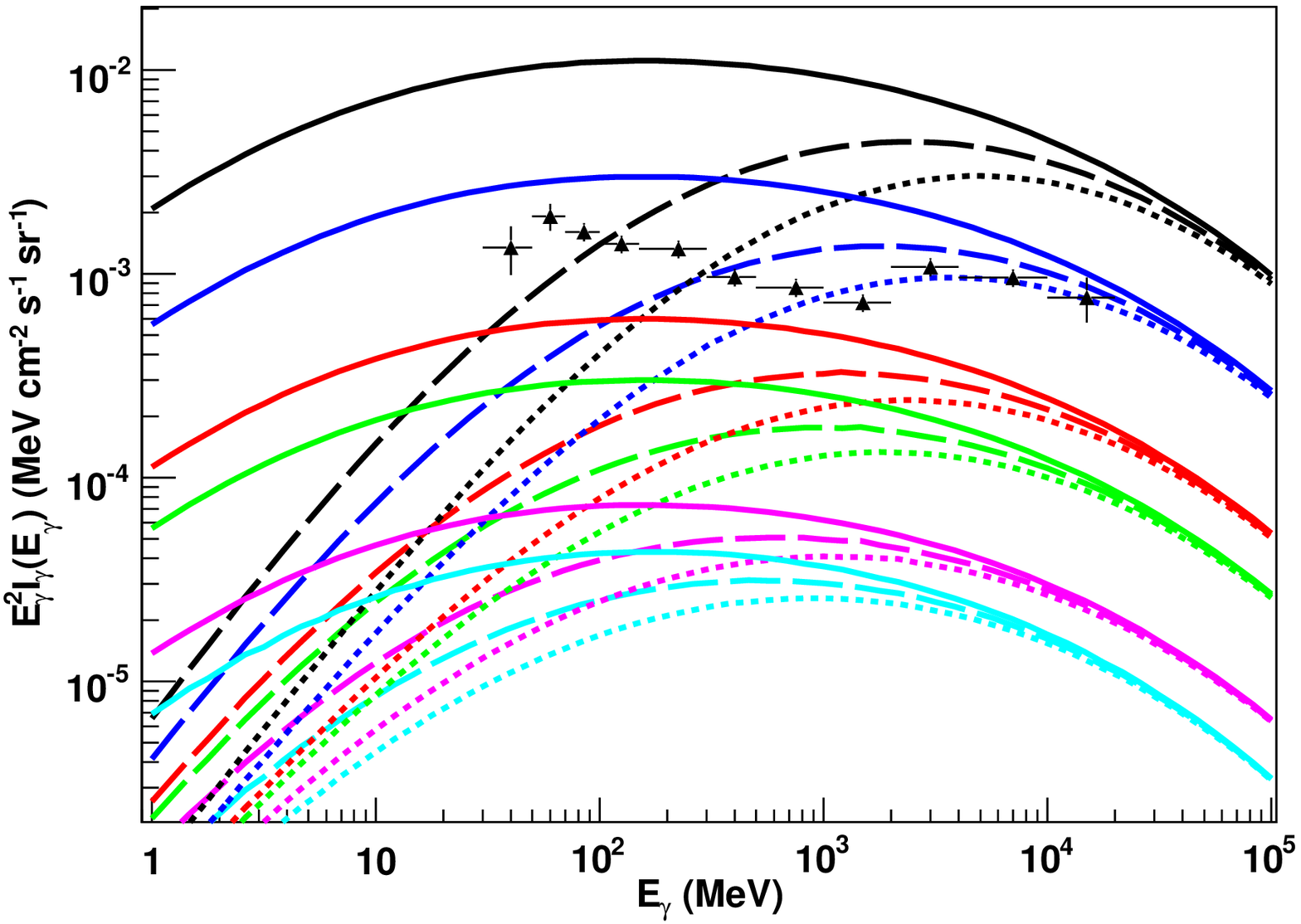}}
\begin{minipage}[tl]{\wcap}
\caption{Integral intensity above 100 MeV
and 1 GeV. {\it Black line,} no modulation; {\it red lines,}
potential $\Phi_1$;  {\it blue lines,}
potential $\Phi_2$. {\it Dashed lines,} $\Phi_0=500$ MV;
{\it dotted lines,} $\Phi_0=1000$ MV.\vspace*{2\baselineskip}}
\label{fig3e}
\end{minipage} \hfill
%
\begin{minipage}[tl]{\wcap}
\caption{Differential intensities for selected $\theta$.
Line-sets (top to bottom): $0.3^\circ$,
1$^\circ$, 5$^\circ$, 10$^\circ$, 45$^\circ$, and 180$^\circ$.
Solid line: no modulation; dashed line: $\Phi_0 = 500$ MV;
dotted line: $\Phi_0 = 1000$ MV.
Data points: diffuse extragalactic \gray\ flux 
(A. W. Strong, I. V. Moskalenko, \& O. Reimer [ApJ, 613,956 (2004)]).
}
\label{fig4e}
\end{minipage} 
\end{figure*}

\begin{table}[h]
\caption{\footnotesize All-sky average integral intensity \label{table1e} \vspace{-2\baselineskip}}
\begin{center}
\begin{tabular}{lcccccc}
\hline
\hline
$E$ && $\Phi_0= 0$ && $500$ MV && $1000$ MV
\\
\hline


$>$10  MeV && 7.1 && 6.5 && 6.0
\\
$>$100 MeV && 1.3 && 1.2 && 1.1
\\
$>$1   GeV && 0.14 && 0.13 && 0.11
\\
\hline
\noalign{\smallskip}
\multicolumn{7}{l}{{\sc Note.} --- Units $10^{-6}$ cm$^{-2}$ s$^{-1}$ sr$^{-1}$.}
\end{tabular}
\end{center}
\end{table}

\end{document}